\documentstyle[12pt,epsf]{article}
\begin{document}
\def \pt {p_{\rm T}}
\def \Et {{\rm E}_{\rm T}}
\def \jpsi {{\rm J}/\psi}
\def \jpsimumu {{\rm J}/\psi \rightarrow \mu^+ \mu^-}
\def \pbarp {{\overline p}p}

\topskip 2cm
\begin{titlepage}

\begin{flushright}
%{\bf CDF/PUB/CDF/3139 \\ FERMILAB-CONF-95/096-E}\\
{\bf FERMILAB-CONF-95/096-E}\\
\end{flushright}

\begin{center}
{\large\bf Charm and Beauty Results} \\
{\large\bf from CDF and D0} \\
\vspace{2.5cm}
{\large Paul Derwent} \\
\vspace{.5cm}
{\sl CDF/University of Michigan }\\
{\sl Ann Arbor, Michigan, 48109 USA}\\
{\sl Reporting for the CDF and D0 collaborations}\\
\vspace{2.5cm}
\vfil
\begin{abstract}

I report, for the CDF and D0 collaborations, results from the 92-93 Tevatron
Collider run concerning charm and beauty quark production and beauty meson
decay properties.
\end{abstract}

\end{center}
\end{titlepage}

\section{Introduction}

\par  Experiments at hadron colliders have become an important
element in the understanding of the physics of heavy flavors (charm and beauty)
in
recent years.  The large production cross sections (on order of 50 $\mu$b for
$b$ quarks at Tevatron energies) and high luminosity of the Tevatron collider
(on order $10^{31}$~cm$^{-2}$~sec$^{-1}$) make plentiful samples.
The CDF and D0 collaborations have exploited this large cross section
through two distinctive trigger signatures, the presence of high transverse
momentum ($\pt$) leptons from the semileptonic decays of the heavy baryons and
the presence of dileptons from the decays of $\jpsi$ mesons.

\par  With these data large samples, production cross section measurements are
made.
These measurements test our current understanding of perturbative QCD, which is
expected to work well for the production of heavy flavors.  In addition,
measurements of the decay properties of the heavy baryons give insight into
the weak mixing angles of the quarks.

\par This paper focuses on results from
the 92-93 Tevatron Collider run.  CDF collected approximately 19 $pb^{-1}$
during this time period and D0 collected approximately 13 $pb^{-1}$.  Detailed
descriptions of the CDF~\cite{CDFNIM} and D0~\cite{D0NIM} detectors can be
found elsewhere.   Topics covered include charmonium production measurements
(from CDF and D0), inclusive and exclusive beauty production  measurements
(from CDF and D0), the $B_S$ meson lifetime (CDF), and the first measurement of
time dependent mixing at a hadron collider (CDF).  Given the large number of
analyses presented, I have not included detail on the experimental issues but
have instead included references where appropriate.   These results, unless
noted otherwise, should be considered preliminary.

\section{Charmonium Production}

\par In $\pbarp$ collisions, $\jpsi$ mesons come from 3 different sources: (1)
direct production, (2) decay of $B$ hadrons, and (3) radiative decay of
$\chi_c$ mesons.  Various methods, described below, are used to disentangle the
three sources, which give information about both charm and beauty production
and fragmentation at low transverse momentum.

\subsection{CDF $\jpsi$ and $\chi_c$ Measurements}

\par
\begin{figure}
\epsfysize=3.5in
\epsffile[0 90 594 684]{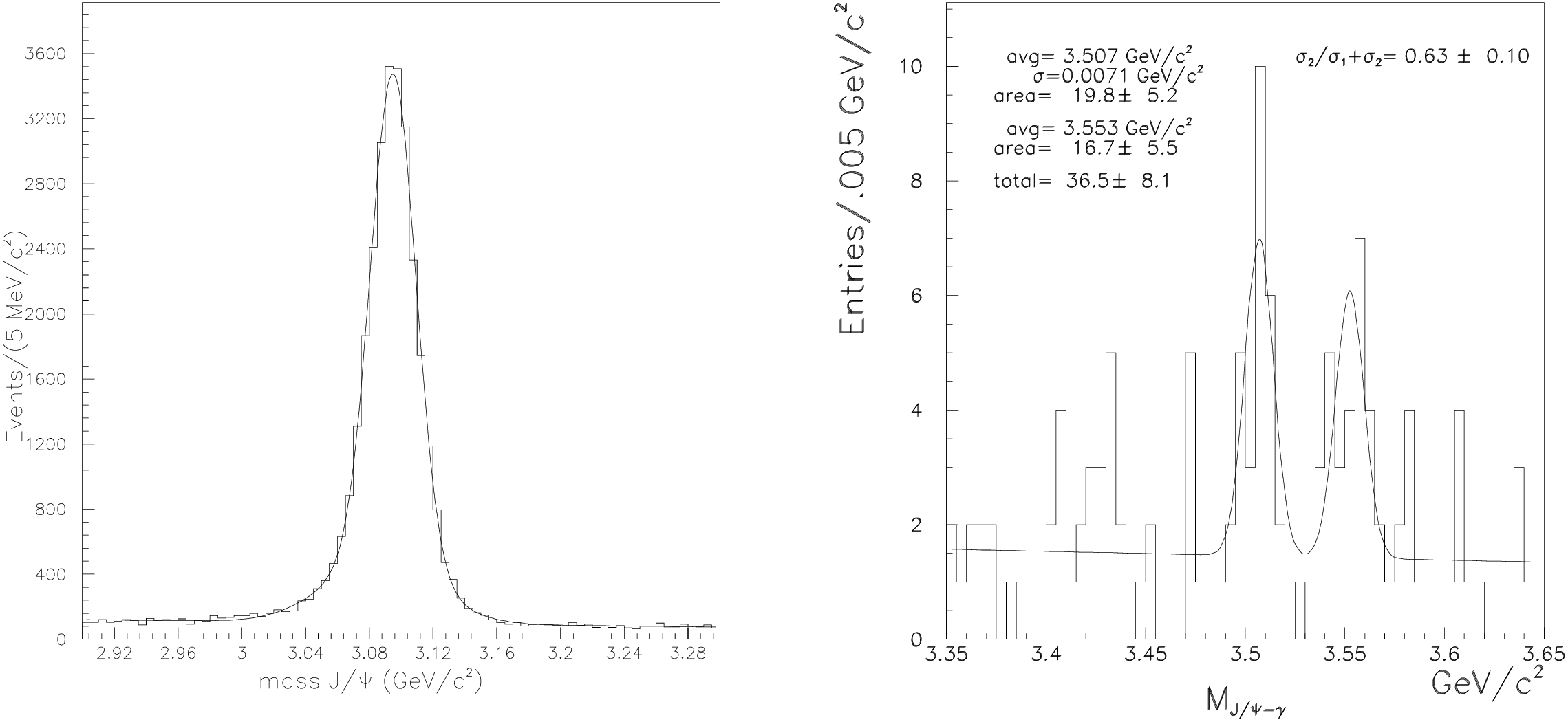}
\caption{The dimuon mass distributions (left) and $\mu \mu \gamma$ (right) mass
distributions from CDF.  The histograms represent the data, while the
smooth curves are the final fits to the data.}
\label{fig-cdf_mass}
\end{figure}

\begin{figure}
\epsfysize=4in
\epsffile[-120 90 474 684]{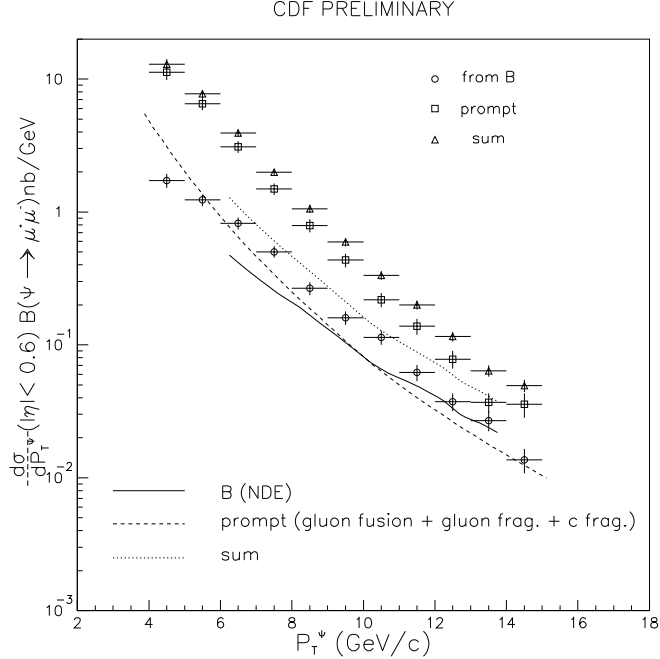}
\caption{The $\jpsi$ differential cross section from CDF, as a function of the
transverse momentum of the $\jpsi$.  Shown separately are the prompt and $B$
hadron components, for both theoretical predictions and experimental
measurement.}
\label{fig-psi_pt}
\end{figure}

 The CDF Collaboration identifies $\jpsi$ mesons through their decay to
$\mu^+\mu^-$ states.  For the measurements presented here, one muon is required
to have $\pt>$ 1.8 GeV/c, and a second muon with $\pt>$ 2.8 GeV/c, with
$\mid\eta^{\mu \mu}\mid <$ 0.6 and $\pt^{\mu \mu}>$ 6 GeV/c.  A fit over the
dimuon mass region 2.9 -- 3.3 GeV/c$^2$ to a Gaussian plus linear background is
used to extract the $\jpsimumu$ component of the dataset (see
figure~\ref{fig-cdf_mass}).  The vertex flight distance is used to measure the
prompt and $B$ hadron fractions~\cite{blife_prl}.  In figure~\ref{fig-psi_pt},
the measured d$\sigma$/d$\pt$ distribution for $\jpsi$ is shown, with the
three sets of data points (and curves) representing the total, prompt
component, and $B$ hadron component of the differential cross section.  A well
defined sample $\psi(2s) \rightarrow \mu^+\mu^-$ has also been identified.
Preliminary differential cross section measurements of the prompt and $B$
hadron of the $\psi(2s)$ are described in reference~\cite{troy_at_dpf}.

\par Radiative $\chi_c$ decays are identified by looking
at the mass difference between $\mu \mu \gamma$ and $\mu \mu$ states.
In a previous publication~\cite{cdf_chi}, CDF has identified
$\jpsi$ from radiative $\chi_c$ decays with the identification of a photon in
the calorimeter.  Results from the current dataset will be available soon.
Another CDF analysis identifies photons which convert to electrons in the
detector
material.  With this technique, the individual $\chi_{c1}$ and $\chi_{c2}$
states can be distinguished, though at much lower efficiency.  In
figure~\ref{fig-cdf_mass}, the $\mu \mu \gamma$ mass, where the photon has been
identified through conversion into an $e^+e^-$ pair, is shown.  With this
technique, CDF has measured the
ratio of $\sigma(\chi_{c2}) / \sigma(\chi_{c1}) + \sigma(\chi_{c2}) = 0.63 \pm
0.10$ (stat) $\pm$ 0.03 (sys) for $\pt^{\mu \mu}>$ 6 GeV/c.

\par With a large sample of $\chi_c$ radiative decays with the
photon identified in the calorimeter, CDF is preparing a measurement of the
fraction of $\chi_c$ from $B$ hadron decays.  This measurement uses the high
precision vertex capabilities to fit the prompt and $B$ hadron components,
similar to what is used for $\jpsi$ and $\psi(2s)$ measurements.

\subsection{D0 $\jpsi$ and $\chi_c$ Measurements}

\par
\begin{figure}
\epsfysize=3.5in
\epsffile[-120 90 474 684]{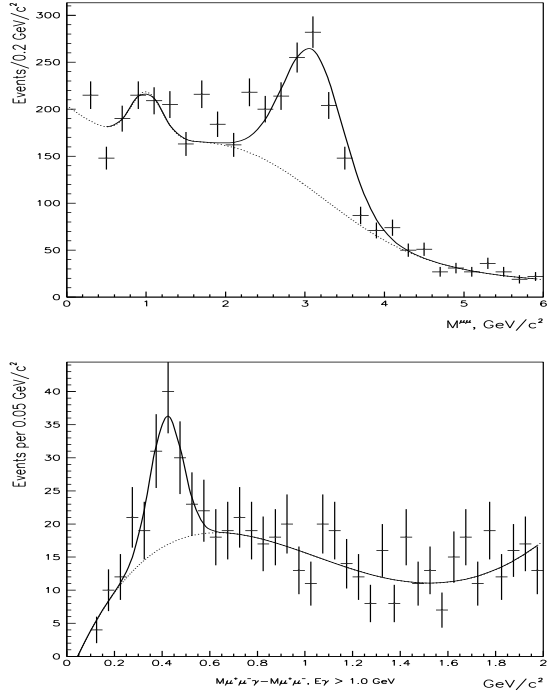}
\caption{The dimuon mass (upper) and $\mu \mu \gamma - \mu \mu$
(lower) mass distributions from D0.  The crosses are the data points,
while the smooth curves are the final fits to the data.  The dotted curves are
the background parameterization.}
\label{figure-d0_mass}
\end{figure}

 The D0 Collaboration also identifies $\jpsi$ mesons through their decay to
$\mu^+\mu^-$ states.  Both muons are required to have
$\pt>$ 3 GeV/c, with $\mid\eta^{\mu \mu}\mid <$ 0.6 and $\pt^{\mu \mu}>$ 8
GeV/c.  In the mass region 0.5 -- 6.0 GeV/c$^2$, a fit to the predicted $\mu^+
\mu^-$ mass distribution from all dimuons sources in $B$ hadron decays, low
mass meson
resonances, Drell Yan, and $\jpsimumu$ gives a total of 444 $\pm$ 36 (stat)
$\pm$ 44 (sys) $\jpsi$ events (see figure~\ref{figure-d0_mass}).

\par  $\jpsi$ from radiative $\chi_c$ decay are identified by measuring the
mass difference between $\mu \mu \gamma$ and $\mu \mu$ (see
figure~\ref{figure-d0_mass}).  Photons are identified
in the calorimeter, with a measured energy greater than 1 GeV.  Both the
$\chi_{c1}$ and $\chi_{c2}$ (which are approximately 45 MeV/c$^2$ apart in
mass) contribute to the identified sample.
The width of the mass difference is dominated by the calorimeter energy
resolution and is measured to be 63 MeV/c$^2$, which doesn't allow for the
resolution of the $\chi_{c1}$ and $\chi_{c2}$ states.  30.5 $\pm$ 7 (stat) $^{+
5.5}_{- 4.4}$(sys) \% of $\jpsi$ with $\pt>8$ GeV/c and $\mid\eta\mid<$ 0.6 are
found to come from radiative $\chi_c$ decays.

\section{Beauty Production}

\par  Initial measurements of beauty quark production at the Tevatron were
consistently higher than next-to-leading order QCD predictions~\cite{89_meas}.
With the higher statistics datasets and improved detectors, measurements from
the 92-93 Tevatron Collider run offer the opportunity for significant
improvements of these measurements.

\subsection{CDF Cross Section Measurements}

\par
\begin{figure}
\epsfysize=4in
\epsffile[-120 90 474 684]{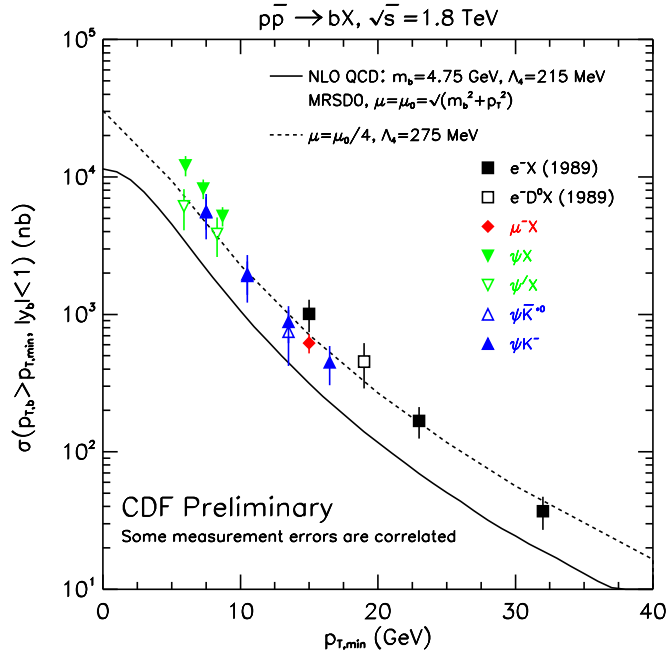}
\caption{The cross sections for $\pt^b>\pt^{min}$ from CDF, using several decay
modes.  The measurements cover a wide range of transverse momentum and are
in reasonable agreement with the upper estimates for the production cross
section.}
\label{fig-cdf_inclusive_92}
\end{figure}

  CDF has used many different decay modes of $B$ hadrons to extrapolate
from the observed cross sections to $b$ quark cross sections, among these the
$B \rightarrow \jpsi X$, $B \rightarrow \psi(2s) X$, $B \rightarrow e X$, $B
\rightarrow \mu X$, $B \rightarrow \jpsi{\overline K}^{*0}$, and $B
\rightarrow \jpsi K^-$.  In
figure~\ref{fig-cdf_inclusive_92}, the most recent measurements of the
inclusive cross section, $\sigma(\pt^b>\pt^{min},\mid y^b \mid < 1)$, with the
standard definition of $\pt^{min}$ of reference~\cite{89_meas}, are shown.  The
experimental measurements are in reasonable agreement with the upper estimates
for the production cross sections.

\par  CDF has also made measurements of differential $\pt^B$ distributions, for
the exclusive decays $B^+ \rightarrow \jpsi K^+$ and $B^0 \rightarrow \jpsi
K^{*0}$, followed by $\jpsimumu$.  The $B$ meson is required to have $\pt>$ 6
GeV/c and $c\tau>$ 100 $\mu$m, with the 3 (4) tracks constrained to come from a
common vertex point.   Under the assumption that $\sigma(B^+) = \sigma(B^0)$, a
simultaneous fit to the $B^+$ and $B^0$ mass distributions is performed, with
the fits constrained by the relative identification efficiencies of the two
resonances (the difference is dominated by the efficiency of finding the low
$\pt$ pion in the $K^{*0}$ decay).

\par
\begin{figure}
\epsfysize=3.5in
\epsffile[0 90 594 684]{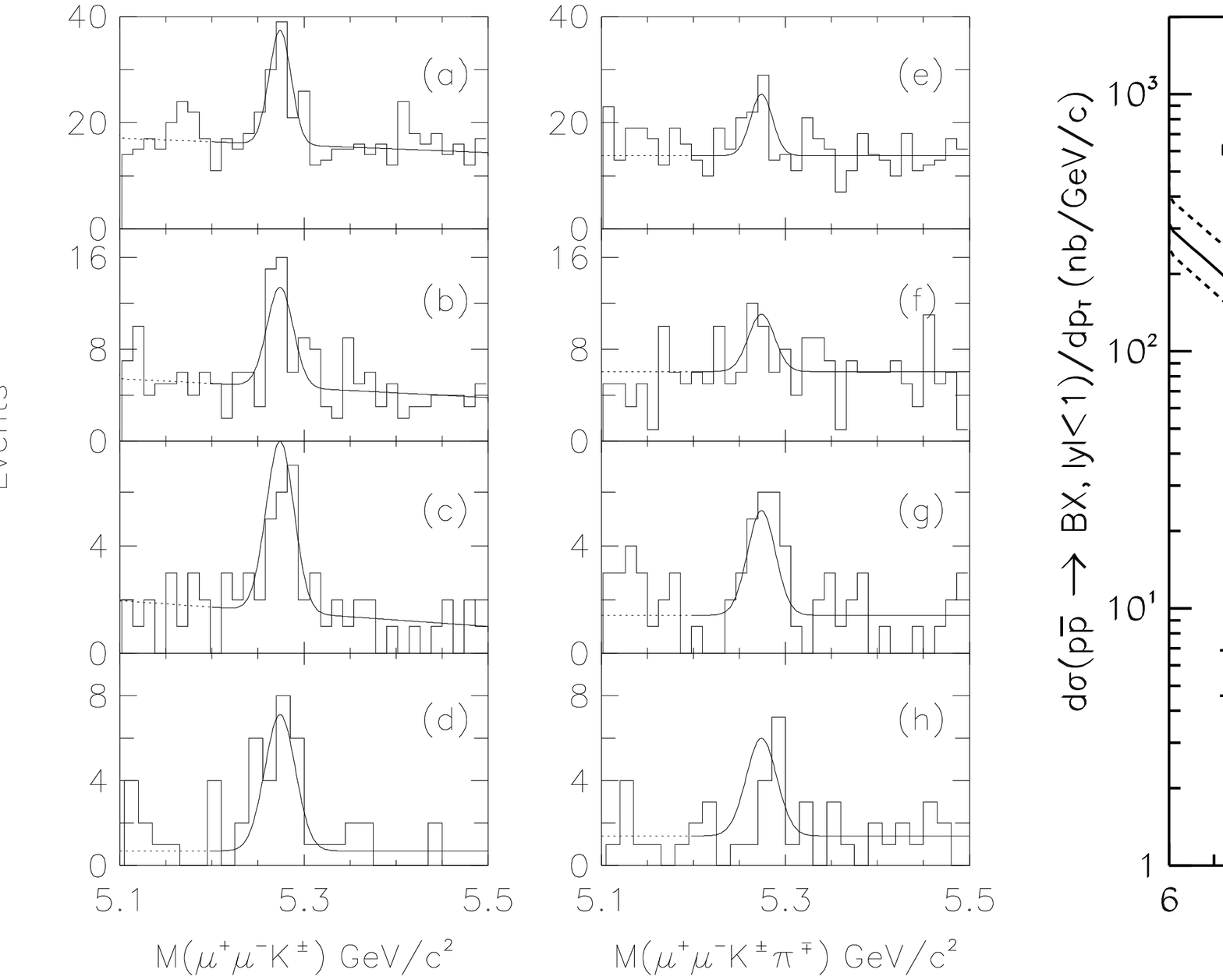}
\caption{On the left, the CDF $B^{\pm}$ and $B^0$ meson invariant mass
distributions for the momentum ranges (a,e) 6 -- 9 GeV/c, (b,f) 9 -- 12 GeV/c
(c,g) 12 -- 15 GeV/c and (d,h) $>$ 15 GeV/c.  On the right, the CDF average $B$
meson differential cross section, with a common systematic uncertainty of
11.9\% shown separately.}
\label{fig-bmesnew_comb}
\end{figure}

 The mass distributions, with accompanying fits, for the 4 $\pt$ bins
considered
and the final d$\sigma$/d$\pt^{B}$ distribution are presented in
figure~\ref{fig-bmesnew_comb}~\cite{excl_b_prl}.  The shape is in reasonable
agreement, but there still exists a normalization difference between the
experimental results and the central value of the prediction.  A fit to the
scale factor gives a value of 1.9 $\pm$ 0.2 (stat) $\pm$ 0.2 (sys), with
$\chi^2$ probability of 20\%.

\par CDF has many additional measurements in preparation for publication, among
them integral and differential correlated $b$ quark cross sections,
and differential $B$ meson cross sections using the semi-leptonic
decays $B \rightarrow l D \nu$, all covering a large range of transverse
momentum.

\subsection{D0 cross section measurements}

\par
\begin{figure}
\epsfysize=3.5in
\epsffile[90 180 584 774]{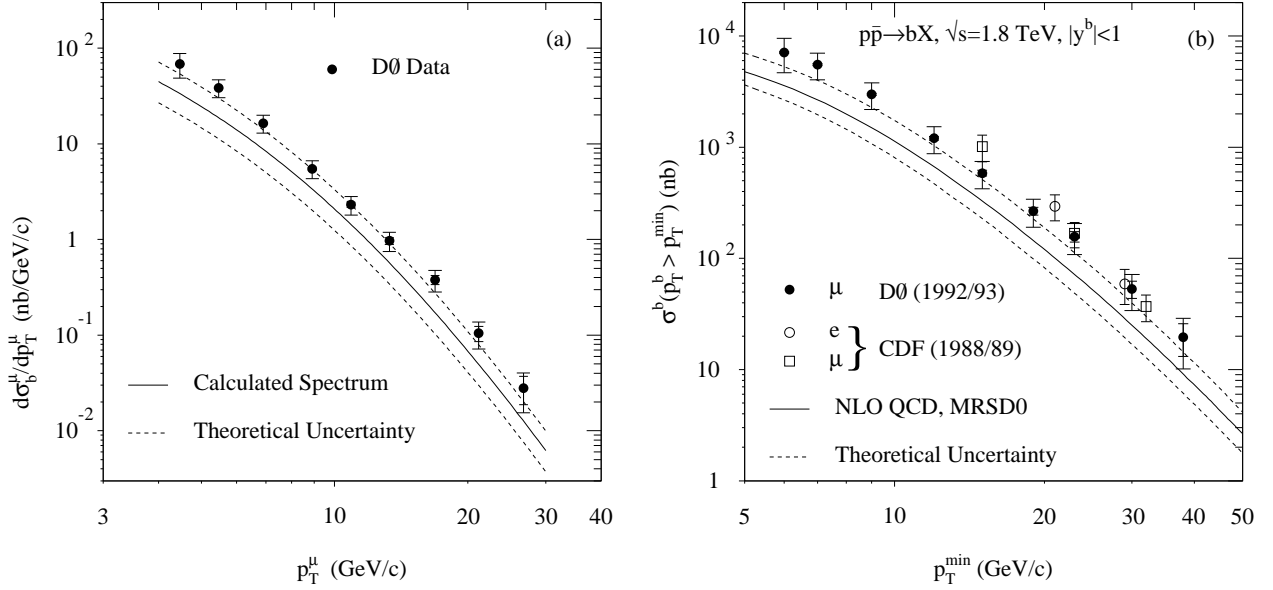}
\caption{The D0 unfolded muon spectrum for inclusive $b$ quark decays (left)
and $b$ quark production cross section (right) compared to NLO QCD predictions.
Inner error bars indicate statistical uncertainties.}
\label{fig-d0_muons}
\end{figure}

  The D0 collaboration has made measurements of the inclusive $b$ quark
cross section using a large inclusive muon sample.  The sample includes
contributions from $b$ decays, $c$ decays, $\pi$ and $K$ decay-in-flight, and
vector boson decays.  For $\mid \eta^{\mu} \mid <$ 0.8, the measured
differential $\pt$ cross section for muons is shown in
figure~\ref{fig-d0_muons}.   An important piece of this measurement is the
proper unfolding of the momentum resolution ($\sigma(1/p)/(1/p) = 0.18 \times
(p-2)/p \oplus 0.008 \times p$), since the inclusive spectrum falls steeply
with
increasing transverse momentum.

\par  Using the distribution of muon momentum relative to the jet axis, D0
extracts the $b$ fraction of the inclusive muon sample as a function of $\pt$.
D0 finds that 60\% of the events in the inclusive sample have an associated jet
which allows the $\pt^{rel}$ measurement and assumes that the $b$ fraction in
the subset applies to the entire sample.  In figure~\ref{fig-d0_muons}, the
differential muon cross section from $b$ decays is presented.  The experimental
measured points are again consistent with the upper estimate of the
next-to-leading order QCD prediction.  This measurement has recently been
published in Physical Review Letters~\cite{d0_mu_prl}.

\par  Using the same technique previously used by UA1~\cite{ua1} and
CDF~\cite{89_meas}, D0 extracts the inclusive $b$ quark cross section for
$\pt^b > \pt^{min}$ from the differential $b \rightarrow \mu$ cross section.
The measurements are in reasonable agreement with CDF measurements and NLO QCD
predictions, though again at the upper estimate of the theoretical predictions.
Figure~\ref{fig-d0_muons} shows the inclusive cross section
$\sigma(\pt^b>\pt^{min},\mid y \mid < 1)$ from the D0 inclusive muon sample.

\section{$B$ Hadron Decay Measurements}

\par  While the production measurements described above mainly test the
predictions of next-to-leading order QCD, $B$ hadron decay measurements test
our understanding of the weak mixing of the quarks.  The
hadron collider experiments are just beginning to explore the large samples
available to make these measurements.

\subsection{$B_s$ Lifetime from CDF}

\par  With the introduction of high precision vertex detectors, CDF has begun
to make significant measurements of the decay properties of $B$ hadrons,
including inclusive $B$ hadron lifetimes~\cite{blife_prl} and exclusive $B$
meson lifetimes~\cite{bmeslife_prl}.  At this conference, CDF presents a new
measurement of the $B_s$ meson lifetime, using the decay mode $B_s \rightarrow
D_s l \nu$, followed by $D_s \rightarrow \phi \pi$, $\phi \rightarrow K^+K^-$.

\par
\begin{figure}
\epsfysize=3.5in
\epsffile[0 90 594 684]{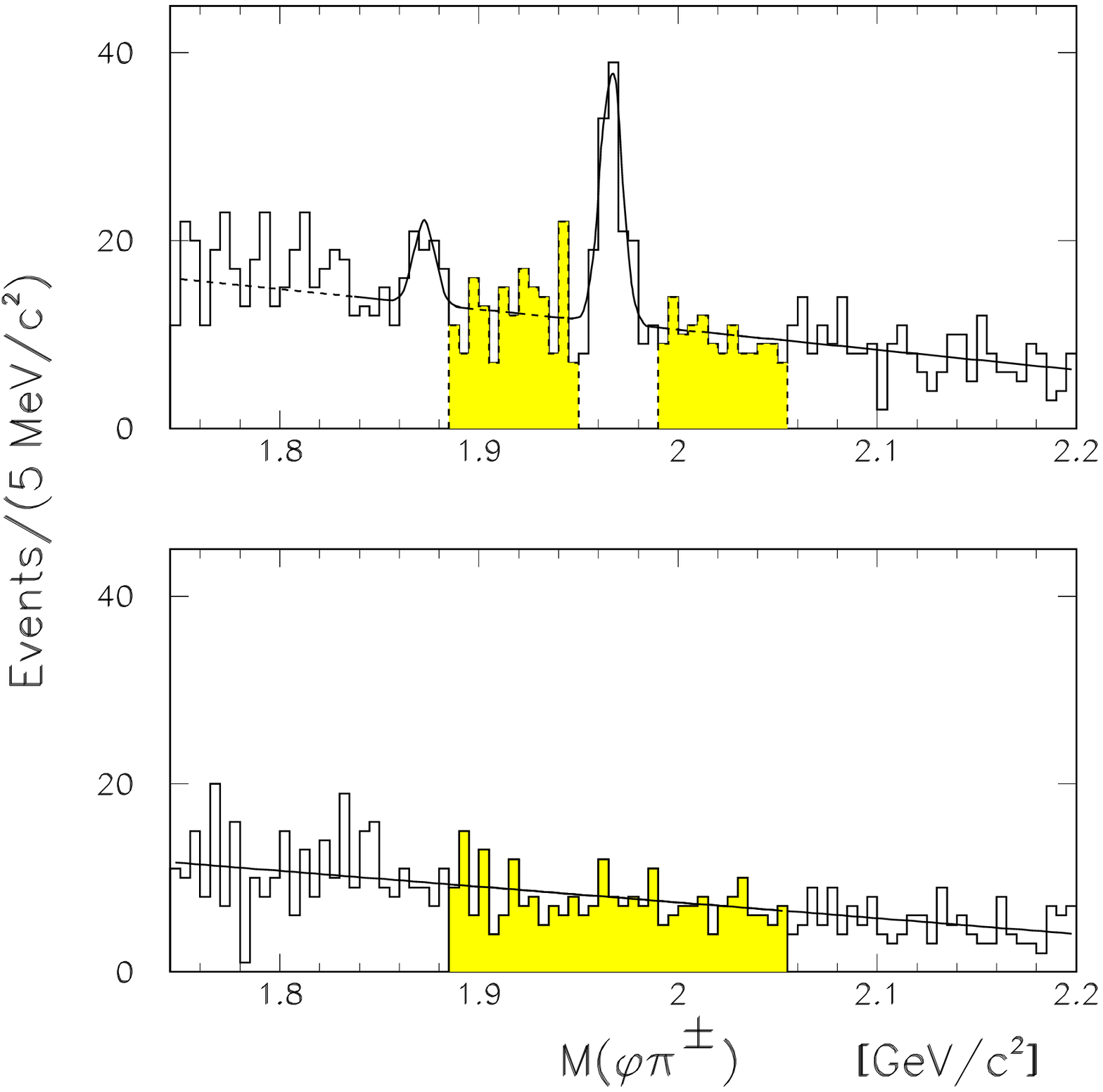}
\caption{On the left, the CDF $\phi\pi$ mass distribution for right sign
combination ($\phi\pi^-l^+$, upper) and wrong sign combination ($\phi\pi^-l^-$,
lower).  On the right, the proper decay length distribution for the $l^+D_s^-$
signal sample, with the contributions from combinatorial background and signal
overlaid.  Inset is the fit to the background sample from the shaded regions in
the mass distributions.}
\label{fig-ds_fits}
\end{figure}

 Using both electron and muon samples, CDF has isolated a significant
sample (76 $\pm$ 8) of $D_s l$ events.  In figure~\ref{fig-ds_fits}, the $\phi
\pi$ mass distribution is displayed.  The upper figure shows the right sign
lepton $+$ pion combination, while the lower figure shows the wrong sign
combination.  Note the presence of the Cabbibo suppressed decay $D \rightarrow
\phi \pi$ on the left in the right sign combination distribution.  Using the
shaded regions to estimate the expected shape of the background sample, an
unbinned likelihood fit to the sum of background plus signal is performed.  The
fit results are also shown in figure~\ref{fig-ds_fits}.  The $D_s$ lifetime is
included in the fit as a free parameter, with resulting measurement of
$c\tau(D_s) = 135^{+40}_{-30} \mu$m (statistical uncertainty only), in good
agreement with the PDG value of 140 $\pm$ 5 $\mu$m~\cite{PDG}.  The final
result is $c\tau(B_s) = 426^{+87}_{-77} \mu$m (statistical $\oplus$ systematic
uncertainties).

\par  Recent theoretical work has suggested that in addition to splitting
between various $B$ meson states, that the CP even and CP odd states of the
$B_s$ meson may also show a lifetime difference on order 10--20\%~\cite{isard}.
$B_s \rightarrow D_s l \nu$ events contain a mix of CP even and CP odd states,
while the sample $B_s \rightarrow \jpsi \phi$ is predominantly CP even.  CDF
has identified a small sample of events (8$^{+3.6}_{-1.6}$) and made a
measurement of the lifetime in this decay mode.  The measurement is currently
statistically limited.  These results have been accepted for publication by
Physical Review Letters~\cite{bslife_prl}.

\subsection{Time Dependent Mixing from CDF}

\par
\begin{figure}
\epsfysize=3.5in
\epsffile[-120 90 474 684]{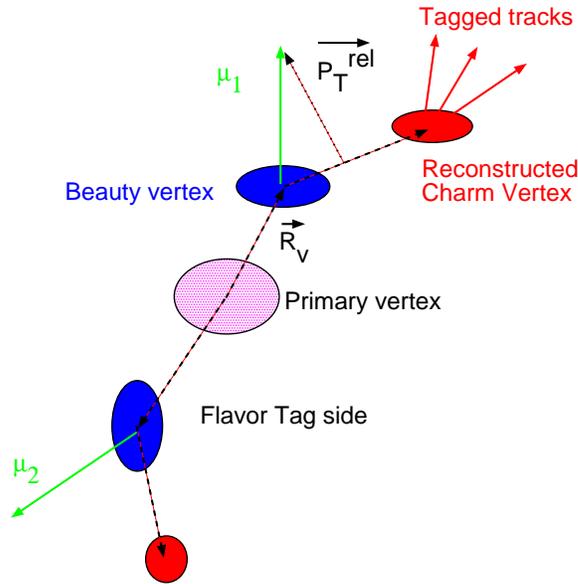}
\caption{A schematic representation of the CDF dimuon mixing analysis.}
\label{fig-mixing_figs}
\end{figure}

 CDF has used a dimuon sample, with a well reconstructed charm vertex, to
measure the time dependent mixing parameter $x_d$.  The combination of a
charm vertex and a muon can be used to reconstruct the proper
decay length of the parent $B$ hadron, where the charge of the muon gives the
flavor of the $B$ hadron at decay time.  The presence of a second well
identified muon gives a second flavor tag.  By measuring the fraction of like
sign muon events as a function of the reconstructed proper decay length, CDF
extracts a measure, $x_d$, of the probability that a $B^0$ meson mixes into
its charge conjugate, ${\overline B}^0$.  Figure~\ref{fig-mixing_figs} shows a
cartoon of an event used in this analysis.

\par  Requiring the presence of two muons with $\pt>$ 2 GeV/c and a
reconstructed charm vertex, CDF finds 1516 events with same
sign muons and 2357 events with opposite sign muons.  The distribution of the
muon $\pt^{rel}$ indicates that the sample is enriched in $B$ decays.
84\% of the muons on the vertex tag side come from primary $B$ decays, while
71\% of the muons on the flavor tag side come from primary $B$ decays and only
12\% are from background.

\par  A binned $\chi^2$ fit to the like sign fraction is used to extract $x_d$,
where the mixing in the $B_s$ sample is assumed to be maximal. CDF constrains
the relative fraction of $B_d$ and $B_s$ mesons to the measured LEP values and
includes the effects of sequential decays in the expected shapes.  The dominant
systematic uncertainty is the understanding of the fraction of events which
come from sequential $B$ decays (e.g., $B \rightarrow D x, ~D \rightarrow \mu
x$).  Detailed studies have shown that the kinematics of the data sample
agree well with the Monte Carlo samples used to generate the expected shapes.

\par
\begin{figure}
\epsfysize=3.5in
\epsffile[-120 -90 474 594]{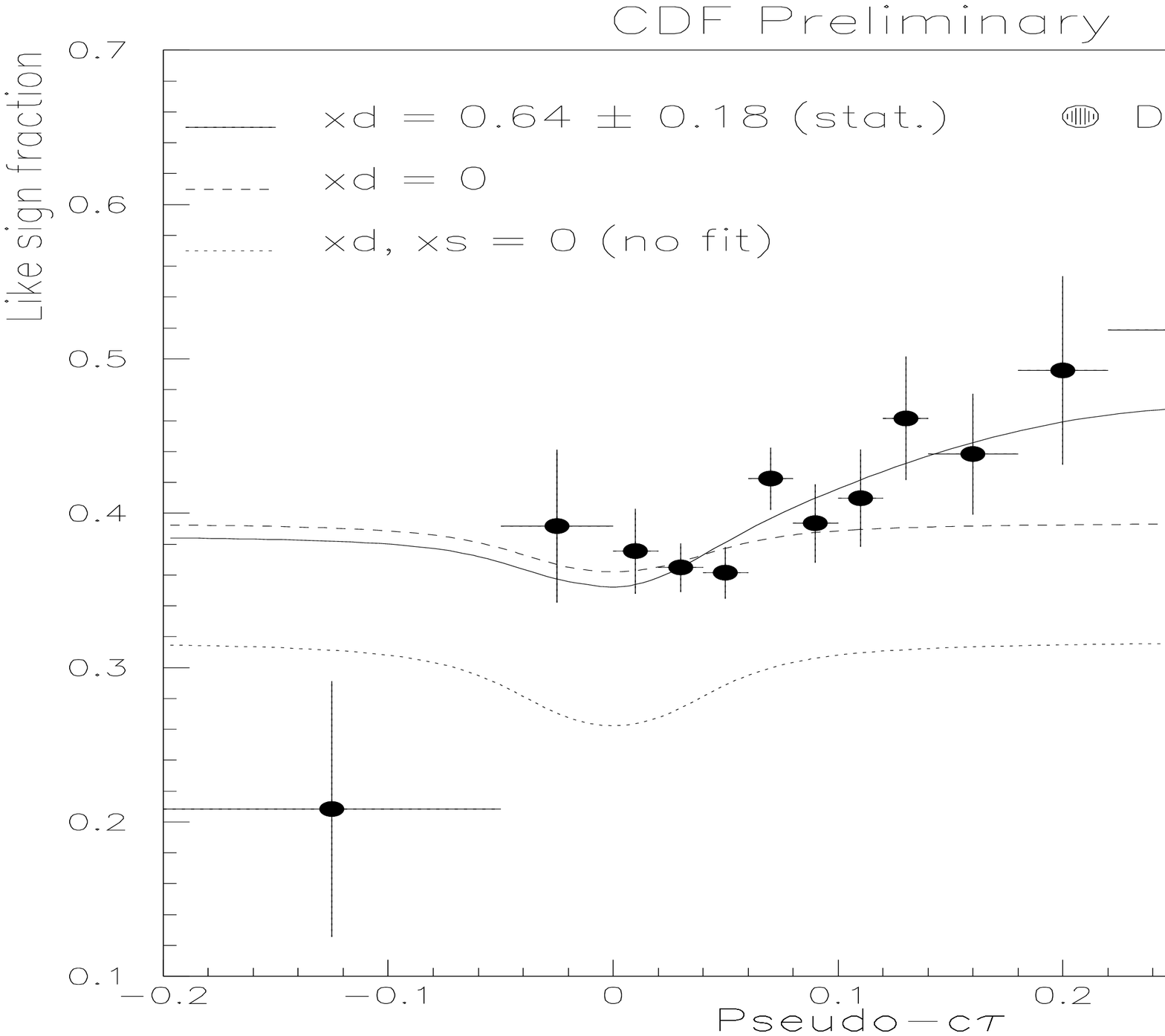}
\caption{The like sign fraction as a function of the
reconstructed proper decay length.  Overlaid are three curves, showing the
effects of the sequential and background contributions, maximal $B_s$ mixing,
and the fit result.}
\label{fig-xd_fit}
\end{figure}

 Figure~\ref{fig-xd_fit} shows the like sign fraction from the data,
overlaid with the fit results under three conditions:  (1)  $x_d$ and $x_s$
are constrained to be 0, (2) $x_d$ is fixed to 0, and (3) $x_d$ is
allowed to float.  The first case shows the effects of the sequential
$B$ decay and background mix in the sample, the second case shows the
effects of maximal $B_s$ mixing, and the third case gives the final fit
results.  The  CDF preliminary result for $x_d$ is 0.64 $\pm$ 0.18 (stat) $\pm$
0.21 (sys).  CDF expects to lower the systematic uncertainty by a factor of 2
in the near future by understanding better the sequential and primary
mix in the sample.

\section{Conclusions}

\par  I have presented recent results from CDF and D0 in charm and beauty
physics.  There are many more results (exclusive and inclusive $B$ hadron
lifetimes,
inclusive and differential cross sections, time integrated mixing results, $B$
decay polarizations) that have not been presented at this
conference.  The Tevatron Collider has been running since January 1994, with
both experiments accumulating significantly larger datasets.  CDF and D0 expect
to decrease the statistical uncertainties in the measurements by factors of 2.5
to 3, depending upon the measurement, with these large datasets.

\section{Acknowledgements}

\par  I wish to thank the members of the CDF and D0 collaborations who
helped me in the preparation of this report.  As a participant in a very
enjoyable and stimulating conference, I would also like to thank Prof.
Belletini and Prof. Greco for the invitation.

\end{document}